\documentstyle{article}

\input{tcilatex}
\begin{document}

\oddsidemargin  -10pt \evensidemargin -10pt \topmargin -47pt \headheight %
12pt \headsep 25pt \footskip 75pt \textheight 9in \textwidth 6.75in %
\columnsep .375in \columnseprule 0pt \twocolumn
\sloppy
\parindent 1em \leftmargini 2em \leftmargin\leftmargini
\leftmarginv .5em \leftmarginvi .5em \flushbottom
\tolerance=1000 
\begin{titlepage}
\begin{flushright}
CFP-IOP-98/01 \\
{\bf January 1998}
\end{flushright}

\vspace{4cm}

\begin{center}
{\Large \bf Unification Picture in Minimal Supersymmetric SU(5) Model \\
with String Remnants}

\bigskip

{\large J.L. Chkareuli and I.G. Gogoladze}

\smallskip

{\it Center for Fundamental Physics, Georgian Physical Society\\
and Institute of Physics, Georgian Academy of Sciences \\
380077 Tbilisi, Georgia}
\end{center}

\bigskip

\bigskip

\begin{abstract}
The significant heavy threshold effect is found in the
supersymmetric $SU(5)$ model with two adjoint scalars, one of
which is interpreted as a massive string mode decoupled
from the lower-energy particle spectra. This threshold related
with the generic mass splitting of the basic adjoint moduli is
shown to alter properly the running of gauge couplings, thus
giving a natural solution to the string-scale grand unification
as prescribed at low energies by LEP precision measurements and
minimal particle content. The further symmetry condition of the
(top-bottom) Yukawa and gauge coupling superunification at a
string scale results in the perfectly working predictions
for the top  and bottom quark masses in the absence of any
large supersymmetric threshold corrections.
\end{abstract}
\end{titlepage}

\twocolumn

Some euphoria caused by the first indication [1] of unification of gauge
couplings extrapolated from their low-energy values under assumption of the
simplest particle content corresponding to the Minimal Supersymmetric
Standard Model (MSSM) seems to be over. In recent years new thorough tests
of the Standard Model have been performed [2]. They show clearly that new
states beyond those of MSSM should be included into play not to be in
conflict with the present precision measurement data, if one contemplated
the gauge coupling unification with not unacceptably high SUSY soft-breaking
scale ($M_{SUSY}>>M_Z$) [3].

On the other hand, the unifying scale $M_U$ at which the couplings would
meet lies one order of magnitude below than the typical string scale $%
M_{STR}\simeq 5\cdot 10^{17}$ GeV [4]. This mismatch, while often considered
as a drawback of string phenomenology, may be interpreted as the most clear
string motivation for a possible grand-unified symmetry beyond MSSM\footnote{%
The matching of those two scales might be qualified as a strong indication
in favor of a pure string unification rather than somewhat superfluous
string-scale Grand Unified Theory (GUT), unless there appears some more
symmetry beyond the GUT at a string scale, e.g. gauge-Yukawa unification,
and the like. We show below that it could be the case.}. In this scenario
[4], the standard gauge couplings would still unify at $M_U$ to form the
single coupling $g_U$, but then it would run up to the string scale $M_{STR}$
and unify with any other gauge ("hidden") and gravitational couplings.

Now, when pondering the possible string-inspired GUT candidates, the minimal
supersymmetric $SU(5)$ model [5] should be considered at the first place,
although it could be stringy realized at higher affine levels $k\geq 2$
only, or in the so-called "diagonal construction" within a $k$-fold product
of the $SU(5)$'s at a level $k=1$ [4]. The $SU(5)$ model with just the light
MSSM particle spectra contributing to the evolution of the gauge couplings
can initiate the additional logarithmic threshold corrections due to the
superheavy states $\Sigma_8(8_c,1)+\Sigma_3(1,3_W)$ and $H_c(3_c,1)+\bar{H}%
_c(\bar{3}_c,1)$, which are left as the non-Goldstone remnants of the
starting Higgs multiplets $\Sigma(24)$ and $H(5)+\bar{H}(\bar 5)$ when $%
SU(5) $ breaks to $SU(3)_C\otimes SU(2)_W\otimes U(1)_Y$ and then to $%
SU(3)_C\otimes U(1)_{EM}$, respectively. At the same time the model has an
additional severe restriction following from proton decay due to the
color-triplet $H_c(\bar{H_c})$ exchange [6]. While at the moment its mass $%
M_c$ can favorably be arranged inside of the narrow interval $2.0\div
2.4\cdot 10^{16}$ GeV (from the negative search for the proton decay and the
detailed renormalization group (RG) analysis for the gauge couplings,
respectively) [7], the model is apparently nearing to be ruled out [8] by a
combination of the improved lower limits on the proton decay mode $p\to\bar{%
\nu}K^+$ from SuperKamiokande and on the superparticle masses at LEP2,
expected in the short run.

In this Letter we argue that the aforementioned adjoint remnants $\Sigma_8$
and $\Sigma_3$, specifically their plausible generic mass splitting, can
play an important part in superhigh-scale physics, thus giving new essential
(yet missing) details to the unification picture in the minimal
supersymmetric ${SU(5)}$ model to overcome the difficulties mentioned above.

It is well known [4,9] that those states appear in many string models as
continuous moduli which is why they can remain relatively light ($%
M_{\Sigma}<<M_{U}$) and, as a result, push the unification scale $M_U$ up to 
$M_{STR}$ [7,9]. At the same time, while a prediction for $\alpha _s$ was
found independent (at one-loop order) of $M_U$ [10] and $M_{\Sigma}$ [7,11],
that is turned out to depend crucially on the mass-splitting between $%
\Sigma_{8}$ and $\Sigma_3$, as it is demonstrated clearly below. Whereas in
the standard one-adjoint superpotential [5] such a splitting is generically
absent at a GUT scale, with a new renormalizable superpotential proposed,
which contains also the second adjoint $\Omega$ (interpreted as a massive
string mode), the large generic mass-splitting appeared between $\Sigma_8$
and $\Sigma_3$ is found fairly ample for gauge coupling unification to be
completely adapted with the present values of $\alpha_{s}(M_Z)$, weak mixing
angle $\theta_{W}$ and top-quark pole mass ($\overline{MS}$ values) [2] 
\begin{eqnarray}
\alpha_{s}(M_Z)=0.119\pm 0.004,~~sin^2\theta_{W}=0.2313\pm 0.0003,  \nonumber
\\
m_t=175.6\pm 5.5~,~~~~~~~~~~~~~~~~  \label{1}
\end{eqnarray}
even though the SUSY soft-breaking scale $M_{SUSY}$ ranges closely to $M_Z$
and the color triplet mass $M_c$ is taken at the GUT scale $M_U$. Both of
cases of the low and high values of $tan\beta$ (the ratio of the vacuum
expectation values (VEVs) of the up-type and down-type Higgs doublets
involved) therewith look in the model to be physically interesting.

We start recalling that, to one-loop order, gauge coupling unification is
given by the three RG equations relating the values of the gauge couplings
at the Z-peak $\alpha _i(M_Z)$ $(i=1,2,3)$, and the common gauge coupling $%
\alpha_U$ [1]: 
\begin{equation}
\alpha^{-1}_i=\alpha^{-1}_U+\sum_p\frac{b^p_i}{2\pi}ln\frac{M_U}{M_P}
\label{2}
\end{equation}
where $b^p_i$ are the three b-factors corresponding to the $SU(5)$ subgroups 
$U(1)$, $SU(2)$ and $SU(3)$, respectively, for the particle lebeled by $p$.
The sum extends over all the contributing particles in the model, and $M_p$
is the mass threshold at which each decouples. All of the SM particles and
also the second Higgs doublet of MSSM are already presented at the starting
scale $M_Z$. Next is assumed to be supersymmetrice threshold associated with
the decoupling of the supersymmetric particles at some single effective
(lumped) scale $M_{SUSY}$ [3]; we propose thereafter the relatively low
values of $M_{SUSY}$, $M_{SUSY}\sim M_{Z}$, to keep sparticle masses
typically in a few hundred GeV region. The superheavy states, such as the
adjoint fragments $\Sigma_8$ and $\Sigma_3$ at the masses $M_8$ and $M_3$,
respectively, and the color-triplets $H_c$ and $\bar{H_c}$ at a mass $M_c$
are also included in the evolution equations (2). As to the superheavy gauge
bosons and their superpartners (X-states), they do not contribute to the
Eqs.(\ref{2}), for they are assumed to lie on the GUT scale $M_U$ ($M_X=M_U$%
), above which all particles fill complete $SU(5)$ multiplets.

Now, by taking the special combination of Eqs.(\ref{2}) we are led to the
simple relation between gauge couplings and the logarithms of the
neighboring threshold mass ratios 
\begin{eqnarray}
12\alpha^{-1}_2-7\alpha^{-1}_3-5\alpha^{-1}_1~=~~~~~~~~~~~~~~~~~~~~~~~~~~~
~~~  \nonumber \\
\frac{3}{2\pi}(-2ln\frac{M_X}{M_c} +ln\frac{M_c}{M_3}-7ln\frac {M_3}{M_8}-%
\frac{19}{6}ln\frac{M_{SUSY}}{M_Z})  \label{3}
\end{eqnarray}
which can be viewed as the basis for giving the qualitative constraints to
the $\alpha _s(M_Z)$ depending on the present (very precise) measurement of $%
sin^2\theta _W$ (\ref{1}) and superheavy mass splitting, when one goes
beyond the MSSM limit ($M_X=M_c=M_3=M_8$). One can see from Eq.(\ref{3})
that $\alpha _s$ increases with $\frac{M_c}{M_3}$ and decreases with $\frac{%
M_X}{M_c}$, $\frac{M_{SUSY}}{M_Z}$ and, especially, with $\frac{M_3}{M_8}$
(the largest coefficient before logarithm). On the other hand, with an
"alive" color triplet $H_c$ ($M_c<M_X$) one can raise the GUT scale $M_X$ by
lowering the masses of the $\Sigma$ remnants (say, $M_3$ with $M_8$ then
calculated) without affecting $\alpha _s$, as theory predicts (to one-loop
order) the entire mass combination $M^2_XM_3$ in addition to $M_c$ and
common gauge coupling $\alpha_U$ (that can be more clearly viewed from the
other running coupling combinations $5\alpha^{-1}_1-3\alpha^{-1}_2-2%
\alpha^{-1}_3$ and $3\alpha^{-1}_2-2\alpha^{-1}_3-\alpha^{-1}_1$ extracted
from Eqs.(\ref{2}), respectively [7,10,11]). When the color triplet is taken
at a GUT scale ($M_c=M_X$), both of masses $M_X$ and $M_3$ and $\alpha_U$%
\footnote{%
As its best, the $\alpha_U$ could be predicted from the string coupling $%
\alpha_{STR}$, if the latter was somehow fixed at the $M_{STR}$ and then
taken to run down to the $M_X$.} are then predicted. And finally, when all
thresholds are taken to be degenerate ($M_c=M_3=M_X$), besides the $M_X$ and 
$\alpha_U$, the strong coupling $\alpha _s(M_Z)$ depending on the generic
mass ratio $M_3/M_8$ can also be predicted -- somewhat naive, while truly
predictive, ansatz of grand unification leading in the standard case ($%
M_3/M_8=1$ at a GUT scale $M_X$) [5] to the unacceptably high values of $%
\alpha _s(M_Z)$ for the physically most interesting SUSY soft-breaking scale
area [3].

However, a somewhat different (more string-motivated) version of the minimal
supersymmetric $SU(5)$ model, we are coming to, with a generically large
mass splitting between $\Sigma_3$ and $\Sigma_8$ suggests an alternative
unification picture. This follows essentially from a general renormalizable
two-adjoint superpotential of $\Sigma$ and $\Omega$ satisfying also the
reflection symmetry ($\Sigma \to -\Sigma$, $\Omega \to \Omega$)\footnote{%
This is assumed to be the gauge type discrete symmetry $Z_2$ inherited from
superstrings [4] and stable under the gravitational corrections.}

\begin{equation}
W=\frac{1}{2}m\Sigma^2+\frac{1}{2}M_P\Omega^2+\frac{1}{2}h\Sigma^2 \Omega+%
\frac{1}{3}\lambda\Omega^3+W^{\prime}  \label{4a}
\end{equation}
where the second adjoint $\Omega$ can be considered as a state originated
from the massive string mode with the (conventianally reduced) Planck mass $%
M_P=(8\pi G_N)^{-1/2}\simeq 2.4\cdot 10^{18}$ GeV, while the basic adjoint $%
\Sigma$ is left (relatively) light when going from the string scale to lower
energies, $m<<M_P$\footnote{%
One could integrate out $\Omega$ and consider the high-order effective
adjoint superpotential of $\Sigma$ only, however it would not look more
instructive for the analysis presented.}. The superpotential includes also
the ordinary Higgs-doublet containing fundamental chiral supermultiplets $H$
and $\bar{H}$ presented in $W^{\prime}$, which will be discussed just below.
It is particularly remarkable that there are no any generically massless
non-trivial string modes, appart from those corresponding to the standard
supersymmetric $SU(5)$ GUT [5].

One can find now from the vanishing F-terms of the adjoints $\Sigma$ and $%
\Omega$ that their basic supersymmetric vacuum configurations, besides the
trivial (symmetry-unbroken) case and the case when the superheavy adjoint $%
\Omega$ alone developes the VEV, include also the desired case when both of
them develop the paralel VEVs which break $SU(5)$ down to $SU(3)\otimes
SU(2)\otimes U(1)$

\begin{eqnarray}
\Sigma=diag(1,~1,~1,~-3/2,~-3/2)\sigma ~,  \nonumber \\
\sigma=\frac{\sqrt{8mM_P}}{h}(1-\frac{m}{M_P}\frac{\lambda }{h})^{1/2}
\simeq \frac{\sqrt{8mM_P}}{h},  \label{5a}
\end{eqnarray}

\begin{equation}
\Omega=diag(1,~1,~1,~-3/2,~-3/2)\omega~,~~~\omega=\frac{2m}{h}  \label{5b}
\end{equation}
with the hierarchically large VEV ratio $r=(2M_P/m)^{1/2}$ inversed to their
masses. After symmetry breaking the non-Goldstone remnants of $\Sigma$ ($%
\Sigma_3$ and $\Sigma_8$) survive, while being a little mixed ($\sim r^{-1}$%
) with $\Omega_3$ and $\Omega_8$, respectively. Remarkably, to the obviously
good approximation ($\frac{h}{\lambda}>>\frac{m}{M_P}$ for any reasonable
values of the couplings $h$ and $\lambda$), already used in Eq.(\ref{5a}),
the (physical) mass ratio of the $\Sigma_3$ and $\Sigma_8$ is definitely
fixed (at a GUT scale) as

\begin{equation}
M_3=10m~,~~~M_8=\frac{5}{2}m~,~~~\frac{M_3}{M_8}=4  \label{6}
\end{equation}
in contrast to $M_3/M_8=1$ in the standard one-adjoint superpotential [5].
Another distinctive feature of the superpotential considered (which can
easily be viewed from Eqs.(\ref{5a}) and (\ref{6})) is the quite moderate
values of the adjoint coupling $h$ even in the case of the pushed
string-scale unification: $h\sim 0.1$ instead of somewhat fine-tuned adjoint
coupling value $\sim 10^{-5}$ in the ordinary one-adjoint case [7,11].

Let us turn now to the electroweak symmetry breaking in the model induced by
the fundamental Higgs supermultiplets $H$ and $\bar H$ in $W^{\prime}$ Eq.(%
\ref{4a}). It is apparent that they must necessarily interact with the basic
adjoint $\Sigma$, since another (superheavy) adjoint $\Omega$ develop too
small VEV (\ref{5b}) to give a reasonable order of mass to the color
triplets $H_c$ and $\bar{H}_c$ after a fine-tuning (to make the accompanied
doublets light) occurs. That is the $\bar{H}H$ pair, along with $\Sigma$,
has to change sign ($\bar{H}H\to -\bar HH$) under reflection symmetry$^3$,
owing to which their mass term can not be included in the superpotential.
Instead, the singlet superfield $S$ ("the 25th component" of the $\Sigma$
with $S\to -S$ defined) should be introduced to make a fine-tuning required.
This part $W^{\prime}$ of the reflection-invariant superpotential Eq.(\ref
{4a}) has a general form

\begin{equation}
W^{\prime}=\bar H(\lambda_1\Sigma+\lambda _2S)H+\frac{1}{2}m^{\prime}S^2
+\lambda^{\prime}S\Sigma\Omega  \label{4b}
\end{equation}
As it can routinely be established from a total superpotential $W$ (\ref{4a},%
\ref{4b}), there always appears possibility in a new supersymmetric vacuum
configuration ($H=\bar H=0$)

\begin{equation}
S=-\frac{\lambda{\prime}}{m^{\prime}}Tr(\Sigma\Omega)\simeq
-15q\sigma~~~(q\equiv\frac{\lambda^{\prime}}{h}/\frac{m^{\prime}}{m})
\label{5c}
\end{equation}
(see $\sigma$ in Eq.(\ref{5a})) to pick a right order of the
coupling-and-mass ratio $q$ so that, from the one hand, not to disturb
noticeably the adjoint vacuum solutions (\ref{5a},\ref{5b}) and masses $M_3$
and $M_8$ (\ref{6}) and, from the other hand, to attain (after a fine-tuning 
$10q=\lambda _1/\lambda _2$) the desired order of the color-triplet mass $%
M_c $ in the vicinity of the unification scale $M_X$, $M_c=\frac{2}{5}\frac{%
\lambda_1}{g_U}M_X$ ($M_X=\frac{5}{\sqrt{2}}g_U\sigma$, $g_U$ is the unified
coupling constant).

So, with the observations made we are ready now to carry out the standard
two-loop analysis (with conversion from $\overline{MS}$ scheme to $\overline{%
DR}$ one included) [1,12] for gauge ($\alpha_1$, $\alpha_2$, $\alpha_3$) and
Yukawa ($\alpha _t$, $\alpha _b$ and $\alpha _{\tau}$ in a self-evident
notation for top- and bottom-quarks and tau-lepton) coupling evolution
depending on, apart from the single-scale ($M_{SUSY}$) supersymmetric
threshold corrections mentioned above, the heavy $\Sigma$ threshold only.
This varies, in turn, from the GUT scale $M_X$ ($M_3=M_X$, $M_8=\frac{1}{4}%
M_X$) down to some intermediate value $O(10^{14})$ GeV pushing thereafter
the $M_X$ up to the string scale $M_{STR}$. The mass splitting between weak
triplet $\Sigma_3$ and color octet $\Sigma_8$ noticeably decreases in
itself, while $M_3$ and $M_8$ run from $M_X$ down to the lower energies, as
it results from their own two-loop RG evolution, which is also included in
the analysis. On the other hand, the color triplets $H_c(\bar H_c)$ are
always taken at $M_X$, for the strings seem to say nothing why any states,
other than the adjoint moduli $\Sigma_3$ and $\Sigma_8$, could left
relatively light.

As to the Yukawa coupling evolution, we consider the two possible cases of
low and large values of $tan\beta $ leading to the proper bottom-tau Yukawa
unification [5,13] with their mass ratio $R=m_{b}/m_{\tau }$ within the
experimental region $R_{exp}(M_{Z})=1.60\pm 0.25$ [2,13] required. The first
case corresponds to the large enough value of $\alpha _{t}$ at a unification
scale $M_{X}$, $\alpha _{t}>0.1$ (while $\alpha _{b}(M_{X})$ is
significantly smaller), evolving rapidly towards its infrared fixed point.
From the observable values of the $t$  quark  mass $m_{t}$ the proper values
of $tan\beta $ are then predicted. The second case, while generally admitted
over the whole area for the starting values for $\alpha _{t}$ and $\alpha
_{b}$ at $M_{X}$, is favorably advanced to the physically well-motivated
(with or without the underlying SO(10) gauge symmetry of no concern)
top-bottom unification case [13] with the relatively low, though still
providing the fixed-point solution, $\alpha _{t}$ and $\alpha _{b}$ values
at $M_{X}$, $\alpha _{t}(M_{X})=\alpha _{b}(M_{X})=0.02\div 0.1$, required.
Here, not only $tan\beta $, but also $m_{t}$ could distinctively be
predicted (from $m_{b}$ and $m_{\tau }$), if there were a more detailed
information about superparticle mass spectrum. Generally the large
supersymmetric loop contributions to the bottom mass are expected which make
uncertain the top mass prediction as well [13] unless the SUSY parameter
sector is arranged in such a way (low values of $\mu $ and $m_{1/2}$ and
large values of $m_{0}$\footnote{%
Those lead to the (relatively) light gluinos and higgsinos and heavy squarks
and sleptons which could be good for the SUSY-inspired proton decay [6-8]
and radiative bottom decay ($b\to s\gamma$) [13], the both enormously
enhanced in the large $\tan\beta$ case. On the other hand, such a
non-uniform superspectrum contradicts, in principle, neither the low SUSY
soft-breaking effective scale $M_{SUSY}$ [3] considered here, nor the
radiative (though being technically unnatural) electroweak symmetry breaking
[13]. While other, less tight, superspectra are also possible [13], we stop
on the above simplest one.}) to make the above radiative corrections to be
negligible [13].

Our results, as appeared after numerical integration of all the RG equations
listed above, are largely summarized in Figs. $1a$ and $1b$. One can see
from them that the $\alpha _s(M_Z)$ values predicted (with a percent
accuracy due to the precise value of $sin^2\theta_W$ (\ref{1}) used and
Yukawa couplings appropriately fixed at $M_X$) are in a good agreement with
the World average value (\ref{1}) for the unification mass $M_X$ ranging
closely to $M_{STR}$.

The values of $\alpha _s(M_Z)$ on the very left of Fig.$1a$ correspond to
the case when $M_3=M_X$ (thresholds are degenerate). This value, $\alpha
_s(M_Z)=0.116\pm 0.001$ (for the $\alpha_t(M_X)=0.3$ taken) , can be
considered as a naive threshold-neglecting prediction of the present model
in the contrast to the analogous value $\alpha _s(M_Z)=0.125\pm 0.001$ in
the standard $SU(5)$ under the same conditions.

In Fig.$1b$ the predicted values of the top quark pole mass are also
exposed. It is readily seen that the experimentally favorable intervals for $%
m_t$ and $\alpha _s$ (1) correspond in much to the same area of $M_X$ in the
vicinity of $M_{STR}$. Interestingly, the unification mass region allowed is
automatically turned out to largely be safe for proton decay through the
Higgs color-triplet exchange [6-8].

Remarkably enough, the presently testable (SUSY threshold neglecting)
top-bottom unification is turned out to work well in the model, thus giving
the good prediction of top-quark mass. Furthermore, the low starting values
of $\alpha _t$ and $\alpha _b$ at $M_X$, as well as the closeness of the
unification mass $M_X$ to the string scale, allow one to make a next step
towards the most symmetrical case which can be realized in the present
string-motivated $SU(5)$ - Yukawa and gauge coupling superunification at a
string scale:

\begin{eqnarray}
\alpha _t(M_X)=\alpha _b(M_X)=\alpha _{\tau}(M_X)=\alpha_U~,  \nonumber \\
M_X=M_{STR}~~~~~~~~~~~~~~~~~~~~~  \label{7}
\end{eqnarray}
This conjecture certainly concerns the third-family Yukawa couplings solely,
since those ones can naturally arise from the basic string-inspired
interactions, whereas masses and mixing of the other families seem to be
caused by some more complex and model-dependent dynamics showing itself at
lower energies. Due to a crucial reduction of a number of the fundamental
parameters the gauge-Yukawa coupling superunification leads immediately to a
series of the very distinctive predictions ( of the $\alpha _s$ in general,
while masses in absence of any large supersymmetric threshold corrections
mentioned):

\begin{eqnarray}
\alpha _{s}(M_{Z}) &=&0.120\pm 0.001~,~~m_{t}=181\pm 1~,  \nonumber \\
\frac{m_{b}}{m_{\tau }}(M_{Z}) &=&1.80\pm 0.01~,~~tan\beta =52\pm 0.2
\label{8}
\end{eqnarray}
in a surprising agreement with experiment [2]. In Fig.2 the superunification
of gauge and Yukawa couplings is demonstrated.

The two concluding remarks concern the further salient features of the
superpotential $W$ proposed.

The first one is that a generic adjoint mass ratio (\ref{6}), while
underlying the self-consistent minimal supersymmetric $SU(5)$ model
presented, remains in a general $SU(N)$ theory broken to $SU(5)$ by the set
of the $N-5$ additional fundamental supermultiplets $\phi^{(k)}$ and $\bar{%
\phi}^{(k)}$ (k=1, ..., N-5), which interact with the adjoint $\Sigma$ via
the general invariant couplings of type $\bar{\phi}\Sigma \phi$ (other terms
are forbidden by the above reflection symmetry$^3$ $\bar{\phi} \phi \to -%
\bar{\phi}\phi$, $\Sigma\to -\Sigma$) included in the superpotential $W$ (%
\ref{4a}). This is to say that one can equally well start from the $SU(N)$
GUT and drive at the same unification picture.

The second one refers to the possible corrections to the superpotential $W$
arising from the high-dimension operators [14] induced at the Planck scale.
Fortunately, due to the same basic reflection symmetry$^3$ of the model,
such operators, if appeared for scalars $\Sigma$ and $S$ developing the
principal VEVs, should have dimension six and higher, $\delta L=\frac{c}{%
M^2_P}~Tr(GG\Sigma^2) + ...$ ($G$ is the gauge field-strengh matrix of the $%
SU(5)$, $c=O(1)$), whose influence on the present model predictions seems to
be negligible in contrast to the standard $SU(5)$ where they can largely be
smeared out [14].

Those and the other (yet applied above) attractive features of the
superpotential $W$ seem to open the way to the natural string-scale grand
unification, as prescribed at low energies by the the gauge coupling
precision measurement and the minimal particle content.

We would like to thank Z.Berezhiani, D. Kazakov, A. Kobakhidze and
K.A.Ter-Martirosyan for helpful discussion and useful comments and
A.Kobakhidze also for drawing our attention to a general $SU(N)$ nature of
the unification picture presented. Financial support by INTAS grant
RFBR-95-567 and Grant of the Georgian Academy of Science 2-10-97 is
gratefully acknowledged.

\newpage

\onecolumn

\noindent {\bf Figure Captions}

\bigskip

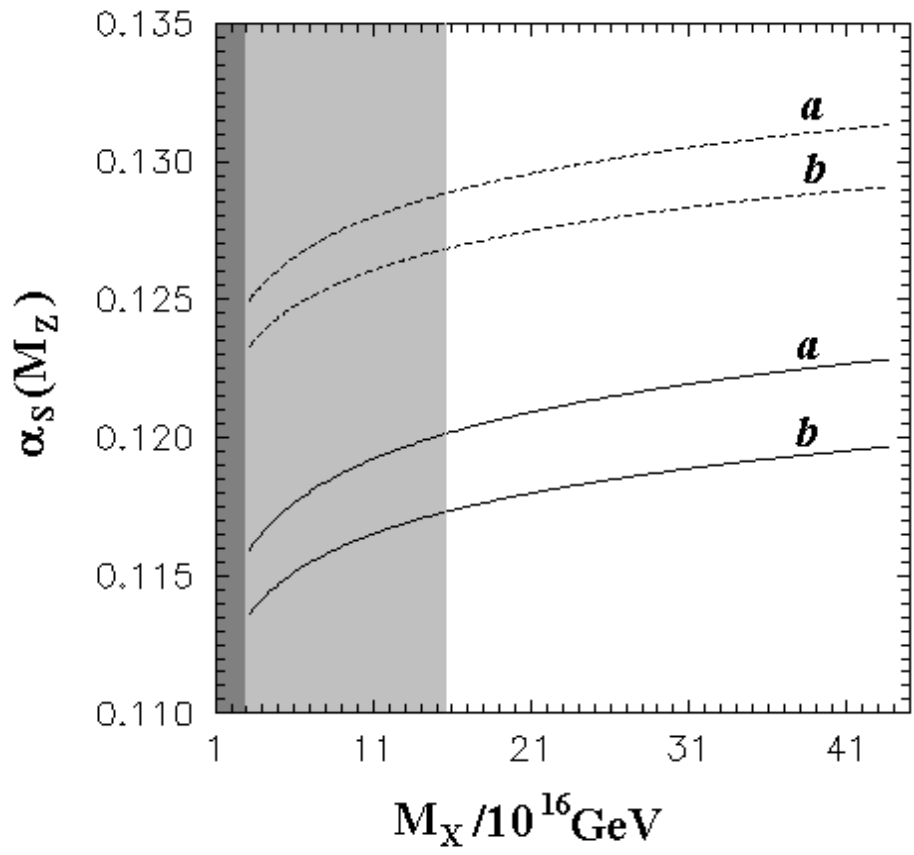
\noindent {\bf Fig.1}~~The predictions in the present model (the solid
lines) and in the standard supersymmetric $SU(5)$ model (the dotted lines)
of $\alpha _{s}(M_{Z})$ as a function of the grand unification scale $M_{X}$
for the two cases of small $tan\beta $ values with top-Yukawa coupling $%
\alpha _{t}(M_{X})=0.3$ taken (a) and large $tan\beta $ values corresponding
to top-bottom unification under $\alpha _{t}(M_{X})=\alpha _{b}(M_{X})=0.05$
(b). In the latter case, the predicted top-quark pole mass values are also
exposed (in the same way) for both of models. The unification mass $M_{X}$
varies from the MSSM unification point ($M_{U}^{0}=10^{16.3}GeV$) to the
string scale ($M_{U}=M_{STR}=10^{17.8}GeV$ for the Kac-Moody level $k=2$;
properly, $M_{\Sigma _{3}}=4M_{\Sigma _{8}}$ $\simeq 10^{14}GeV$ is taken at
a string scale), while the color-triplet mass is assumed to be at
unification scale in all cases ($M_{c}=M_{X}$). The all-shaded areas on the
left of the figures (a) and (b) are generally disallowed by the present
bound [2] on nucleon stability.

\bigskip

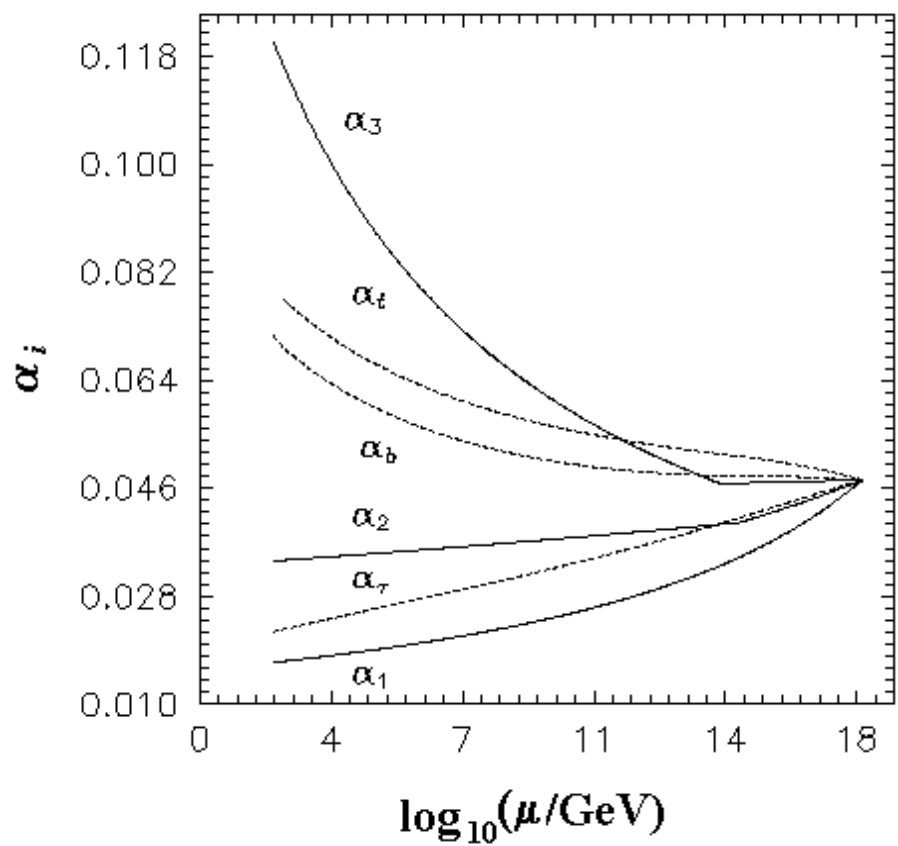
\noindent {\bf Fig.2}~~The superunification of gauge ($\alpha _1$, $\alpha
_2 $, $\alpha _3$) and Yukawa (top, bottom, tau) couplings at the string
scale (the solid and dotted lines, respectively).

\end{document}